\documentstyle[preprint,prl,aps]{revtex}
\draft
\begin{document}
\title{On the universality of a class of annihilation-coagulation models.}
\author{Daniele Balboni\cite{ack_bal,ack_uno}, Pierre-Antoine Rey\cite{ack_uno}
and Michel Droz \cite{ack_uno}}
\address{D\'epartement de Physique Th\'eorique, Universit\'e de Gen\`eve,
CH-1211 Gen\`eve 4, Switzerland.}
\author{\bf UGVA-DPT 1995/05-892}
\date{\today}
\maketitle
\begin{abstract}
A class of $d$-dimensional reaction-diffusion models interpolating continuously
between the diffusion-coagulation and the diffusion-annihilation models
is introduced. Exact relations among the observables of different models
are established. For the one-dimensional case, it is shown how correlations
in the initial state can lead to non-universal amplitudes for time-dependent
particles density.

\vspace {0.3truecm}
\end{abstract}
\vfill\eject
\section{Introduction}
In the recent years, reaction-diffusion problems have stimulated a large
body of work in many different directions~\cite{RITTEN,LEE}.
The simplest example is provided by the annihilation process $A+A
\rightarrow 0$, in which the $A$ particles diffuse  and react by pairs
on contact.

Usually such kinetic processes are described in terms of macroscopic
rate equations giving the time evolution for the local averaged
concentrations.  One assumes that the reaction
is completely described in terms of the local average densities, and that the
reaction introduces no correlations between the reacting species.  This
is reminiscent of a mean-field like approximation in statistical
physics. However, an important aspect of the problem is neglected,
namely the microscopic fluctuations and it is well known  that these
fluctuations play an important role in low dimensional
systems~\cite{WIL,LEBO,CCD}.

Indeed, for the annihilation process $A+A \rightarrow 0$, the rate
equation predicts that, in the long time regime, the concentration of
$A$ will decrease in time as $a \simeq {\cal A} t^{-1}$ while a
calculation taking into account the local microscopic fluctuations in the
particle density  gives~\cite{KANG} $a \simeq {\cal A} t^{\alpha}$,
with $\alpha={\rm Min}(1,\frac{d}{2})$, where $d$ is the dimensionality
of the system. Moreover, for a Poissonian initial state, the amplitude
${\cal A}$ is a universal quantity, which is in particular
independent on the initial density of the particles.  This nonclassical
power law behavior is called anomalous kinetics.

Thus to describe correctly these systems, it is crucial to work within
a formalism which is able to  keep track of the fluctuations. This is
a difficult task and this is the reason why exact analytical results are
scarce, but for one-dimensional systems. Complex reaction-diffusion systems
are present in nature~\cite{BORN}. However, the study of simple models,
like the diffusion-coagulation process (DC) $A+A \rightarrow A$, or the
diffusion-annihilation one (DA) $A+A \rightarrow 0$ provide a very useful
testing ground for new theoretical approaches.

Different methods have been explicitly developed for the one-dimensional
systems (see for example~\cite{RITTEN,RACZ,LUSH,BALD,SPOU,PRIV,GRYN}).
In most of them, one maps the reaction diffusion problem onto a different
model (quantum chain~\cite{RITTEN}, kinetic Ising model~\cite{RACZ},
invasion process~\cite{BALD}, etc), which is exactly solvable. A different
approach consists in finding the "good observables" for which an exact
equation of motion can be derived and solved. For the DC model, Doering
{\it et al}.~\cite{LIN} have shown that the system was best analyzed in
terms of the time dependent probability ${\cal E}(x,y,t)$ that the interval
between $y$ and $x$ is empty at time $t$. The particle density $a(x,t)$
is simply given by $a(x,t)=-\frac{\partial}{\partial x}{\cal E}(x,y,t)
\vert_{x=y}$. In the continuous limit, ${\cal E}(x,y,t)$ obeys to a simple
diffusion  equation which can be solved analytically. However, Doering
{\it et al.}\/\ were not able to find a similar quantity for the DA model.

In arbitrary dimension, it was soon recognized, based on dynamical
renormalization group arguments~\cite{PELITI_UNO,SAS}, that the DA and
DC models belong to the same universality class. The power law exponents
describing the decay of the particle densities are the same.
However, for the amplitude $\cal A$, the situation turns out to be more
subtle. Whereas for the DC model this amplitude seems to be strictly
independent of the initial state, this is no longer true for the
one-dimensional DA model. Indeed, as noticed by Family and Amar~\cite{FAMILY},
short-ranged  initial correlations can change the asymptotic amplitude in a
continuous way. In their analysis, the DA model is mapped onto a kinetic Ising
model at zero temperature~\cite{RACZ}. Explicit asymptotic results for the
average domain size, average magnetization squared and pair-correlation
function are derived for arbitrary initial conditions. For the case of an
initial magnetization $m_0=0$, the results for the DA model with Poissonian
initial conditions are reproduced. However, for $m_0 \not = 0$, the
particle density has a nontrivial dependence on the initial magnetization.
A nonzero value of $m_0$ means for the DA model that there are some
correlations among the particles in the initial state. What is the mechanism
responsible for this dependence and could a similar situation occur also in
the DC process for which there is no mapping onto a kinetic Ising model?

To answer this question we shall, in this paper, revisit the problem of
universality in the DC and DA model. The strategy is to study the
field theory which can be associated to a reaction-diffusion process,
following the method introduced by  Doi~\cite{DOI} and revisited by
Grassberger {\it et al}.~\cite{GRASS} and Peliti~\cite{PELITI_DUE}.
We shall consider a class of diffusion-coagulation-annihilation models
(called $\alpha$-models), allowing us to go continuously from the DC
process to the DA one.

The paper is organized as follows. In Section 2, the way to associate a field
theory to a reaction-diffusion process is briefly reviewed and the
$\alpha$-models are introduced. Within a functional integral formalism,
different $d$-dimensional $\alpha$-models are related one to the others.
In Section 3, the one-dimensional DC model is revisited. It is shown how the
equation for the time dependent probability ${\cal E}(x,y,t)$ that an
interval between $y$ and $x$ is empty at time $t$ can be obtained within the
field theoretical framework. In Section 4, we introduce the quantity
${\cal T}^{(\alpha)}(x,y,t)$ which is the natural generalization for the
$\alpha$-model of ${\cal E}(x,y,t)$. Its equation of motion can be derived
and solved exactly, leading to an exact expression for the asymptotic behavior
of the particle density. It is shown explicitly how and when the amplitude
universality can be violated. In particular, it is proved that for an initial
state in which particles are pairwise correlated, the violation of amplitude
universality is only possible for the DA model. Finally, some extensions of
this work are discussed in the conclusion.

\section{Field theoretical approach}

For the sake of completeness and to set up the notations,
we briefly sketch the main steps necessary to derive the field theory
associated to a d-dimensional  reaction-diffusion process. We follow the
method introduced by Doi~\cite{DOI} and revisited by Grassberger
{\it et al}.~\cite{GRASS} and Peliti~\cite{PELITI_DUE}.

We start from a description of the system in terms of the probabilities
$P_n(x_1,\dots,x_n;t)$ that a configuration with $n$ particles at points
$x_1,\dots, x_n$ is realized at time $t$. As we are not interested in
specifying which particle is where, we choose $P_n(x_1,\dots,x_n;t)$ to be
a symmetric function in the variables $x_1,\dots, x_n$. Note that, as the
particle number is not conserved, a state of the system  must be specified
by the entire set $\Phi(t)=\{P_n (t)\}_{n=0,\dots,\infty}$ with the
normalization:
$$\sum_n {1\over n!}\int dx_1\dots dx_n \ P_n(x_1,\dots,x_n;t)=1\quad .$$
The dynamics is defined by the master equation:
  \begin{equation}
\partial_t P_n(x_1,\dots,x_n;t) = [{\cal H} P]_n(x_1,\dots,x_n;t)\quad ,
  \end{equation}
where ${\cal H}$ is an operator acting on the set of the
probabilities~\cite{DOI,GRASS}.

Let us consider as an example the following DA process. The $A$ particles
diffuse
and annihilate each other with a reaction rate $V(\vert x-y\vert)$ depending
on the distance. The corresponding master equation reads:
  \begin{equation}
\partial_t P_n(x_1,\dots,x_n;t)=
D\sum_{i=1}^{n}{\partial^2 \over \partial x_i^2}P_n(x_1,\dots,x_n;t)
  \label{equ:master}\end{equation}
$$-{1\over 2}\sum_{i\ne j}V(\vert x_i-x_j\vert)P_n(x_1,\dots,x_n;t)
+{1\over 2}\int dydz\ V(\vert y-z\vert)P_{n+2}(y,z,x_1,\dots,x_n;t)\quad ,$$
where $D$ is the diffusion constant.

We introduce a Fock space representation for the states of the system.
Space-valued annihilation $\psi(x)$ and creation $\pi(x)=\psi^{\dag}(x)$
operators are introduced. The vacuum state $\vert0\rangle$ is defined by
$$\psi(x)\vert0\rangle =0$$
and the annihilation and creation operators obey bosonic commutation
relations:
$$[\psi(x),\pi(y)]=\delta(x-y)\quad .$$
The state specified by the set $\Phi=\{P_n\}_{n=0,\dots,\infty}$ reads:
  \begin{equation}
\vert \Phi\rangle =\sum_{n=0}^{\infty}{1\over n!}\int dx_1\dots dx_n\
P_n(x_1,\dots,x_n)\pi(x_1)\dots\pi(x_n)\vert 0\rangle
  \label{equ:vettore}\end{equation}
and the statistical average $
\langle A \rangle$ over the state $\Phi$ of an observable
$\{A_n(x_1,\dots,x_n)\}_{n=0,\dots,\infty}$ defined by
$$\langle A\rangle=\sum_n {1\over n!}\int dx_1\dots dx_n\
A_n(x_1,\dots,x_n)P_n(x_1,\dots,x_n)$$
takes the form of a scalar product:
$$\langle A\rangle=\langle\vert A\vert \Phi\rangle\quad ,$$
where $A$ is the corresponding operator of the Fock space, such that
$\langle 0\vert \psi(x_1)\dots\psi(x_n)A\vert \Phi \rangle=A_n(x_1,\dots,x_n)$
and $\langle\vert$ is a projection state given by
$$\langle \vert=\langle 0\vert e^{\int dz\ \psi(z)}\quad ,$$
with the property $\langle\vert\pi(x)=\langle\vert$. In particular,
the particle density is given by  $c(x)=\langle \psi(x)\rangle$ and
the n-point correlation functions by $\rho(x_1,\dots,x_n)=\langle
\psi(x_1)\dots\psi(x_n)\rangle$. The normalization of the state is expressed
by the condition $\langle\vert \Phi \rangle=1$. A correlated state,
characterized by the $p$-points cumulants:
$$\langle \psi(x_1)\dots\psi(x_p)\rangle_{cum}=g_{(p)}(x_1,\dots,x_p)$$
has the form:
  \begin{equation}
\vert \Phi \rangle= \exp \left[
\sum_{p=1}^{\infty}{1 \over p!}\int dx_1 \dots dx_p\
g_{(p)}(x_1 ,\dots,x_p)(\pi(x_1)-1)\dots(\pi(x_p)-1)
\right]\vert 0\rangle\quad .
  \label{equ:state}\end{equation}

It can be shown that the master equation can be written in
the Schr\"odinger-like form:
  \begin{equation}
{\partial \over \partial t}\vert \Phi(t)\rangle =H\vert \Phi(t)\rangle\quad .
  \label{equ:schroedinger}\end{equation}
For our annihilation model,  the non-Hermitian evolution operator $H$
is given by
$$H=H_0+H_a\quad ,$$
with
  \begin{equation}
H_0=D\int dz\ \pi(z)\nabla^2\psi(z)
  \label{equ:H_0}\end{equation}
and
  \begin{equation}
H_a=-\int dzdz^{\prime}\ V_a(\vert z-z^{\prime}\vert)[\pi(z)\pi(z^{\prime})-1]
\psi(z)\psi(z^{\prime})\quad .
  \label{equ:H_a}\end{equation}
A second example is provided by the diffusion-coagulation   process.
The interaction part of the evolution operator becomes:
  \begin{equation}
H_c =-\int dzdz^{\prime}\ V_c(\vert z-z^{\prime}\vert)[\pi(z)\pi(z^{\prime})-
\pi(z)]\psi(z)\psi(z^{\prime})
  \label{equ:H_c}\end{equation}
If now both coagulation and annihilation  reactions are simultaneously allowed,
one has:
 \begin{equation}
H_{a+c}=H_a +H_c =
-\int dzdz^{\prime}\ [(V_a +V_c)\pi(z)\pi(z^{\prime})-
V_c\pi(z)-V_a]\psi(z)\psi(z^{\prime})\quad .
  \end{equation}
Although not Hermitian, these evolution operators have the property to
preserve the state normalization: $\langle\vert H=0$. The evolution equation
of an observable $A$ is:
  \begin{equation}
{\partial\over\partial t}\langle A\rangle=\langle AH\rangle=\langle
[A,H]\rangle
  \quad .\label{equ:evoluzioneA}\end{equation}

It can be shown~\cite{PELITI_DUE}, that the correlation functions may
be expressed by functional integrals as:
  \begin{equation}
\langle \psi(x_1,t_1)\dots \psi(x_n,t_n)\rangle _{{\cal S},F_0} =
\int {\cal D}\psi {\cal D}{\bar \psi}\ \psi(x_1,t_1)\dots \psi(x_n,t_n)
e^{-{\cal S}} F_0 [{\bar \psi} (x,0)]\quad .
  \label{equ:integralefunzionale}\end{equation}
The action ${\cal S}$ appearing in the measure $e^{-\cal S}$ is related to
$H$ as follows:
$${\cal S}[{\bar\psi},\psi]
=\int dtdx\ {\bar\psi}\dot{\psi}-\int dt\ H[\pi={\bar\psi}
+1,\psi]\quad .$$
$F_0$ is a functional of the auxiliary field ${\bar \psi}$ at t=0,
and is related to the cumulants of the initial state by
  \begin{equation}
F_0[{\bar \psi}]=
{\rm exp}\left[\sum_{p=1}^{\infty}{1 \over p!}\int dx_1 \dots dx_p\
g_{(p)}(x_1 ,\dots,x_p;0){\bar \psi}(x_1)\dots
{\bar \psi}(x_p)\right]\quad .
  \label{equ:state2}\end{equation}

Let us now assume that $V_c$ is proportional to $V_a$. This leads us to
introduce
a class of model depending on a continuous real parameter $\alpha$ (called
the $\alpha$-models) and defined by the following action:
  \begin{equation}
{\cal S}_{\alpha}=
\int dtdz\ {\bar\psi}\left({\dot\psi}-D\nabla^2\psi\right)+
\int dtdzdz^{\prime}\ V(\vert z-z^{\prime}\vert)
[{\bar\psi}(z){\bar\psi}(z^{\prime})+\alpha{\bar\psi}(z)]
\psi (z)\psi(z^{\prime})\quad .
  \label{equ:Sac}\end{equation}
The simple coagulation and annihilation reactions correspond respectively to
$\alpha=1$ and $\alpha=2$ and will be labeled by the indices $c$ and $a$ in
the following. Thus the $\alpha$-models ($\alpha\in[1,2]$) interpolate
between the coagulation process and the annihilation one.

We are now in the position to relate dynamics with different values of $\alpha$
in a very simple way. Taking the coagulation case as reference,
eq.~(\ref{equ:Sac}) gives:
$${\cal S}_{\alpha} [{\bar \psi}, \psi]
={\cal S}_c [{{\bar \psi} \over \alpha},\alpha \psi]$$
and, by simply rescaling the functions in the functional integral, we get
the following relations between correlation functions:
  \begin{equation}
\langle \psi(x_1,t_1)\dots \psi(x_n,t_n)\rangle _{\alpha,F_0}
=\alpha^{-n} \langle \psi(x_1,t_1)\dots \psi(x_n,t_n)\rangle
_{c,F_0 ^{(1/\alpha)}}\quad ,
  \label{equ:equiv}\end{equation}
where:
$$F_0 ^{(1/\alpha)}[{\bar \psi}]
=F_0 [\alpha{\bar \psi}]\quad .$$
In particular this means that, if at $t=0$ we consider two states
$\vert \Phi_{\alpha}(0)\rangle$ and $\vert \Phi_c(0)\rangle$ such that:
$$\langle\vert \psi(x_1)\dots \psi(x_n)\vert \Phi_{\alpha}(0)\rangle
=\alpha^{-n} \langle\vert \psi(x_1)\dots \psi(x_n)\vert \Phi_c(0)\rangle
\quad ,$$
for all $n$, then this relation in conserved for all times, if
$\vert \Phi_{\alpha} \rangle$
evolves according to the dynamic of the $\alpha$-model and $ \vert \Phi_c
\rangle$ according
to the dynamic of the coagulation one. Moreover, if a set of equations
among correlation functions, or more generally among operators, is satisfied
for the coagulation case, then a similar set of relations exists for the
$\alpha$-model, providing that one rescales the fields according to
$\psi\rightarrow \alpha\psi$. Note that similar relations have been recently
derived in a different context by Henkel {\it et al}.~\cite{MALTE}.

For the particular case $\alpha=2$ and for homogeneous Poissonian initial
conditions for both $\vert \Phi_c \rangle$ and $\vert \Phi_a \rangle$, we have
the following relation
between the concentrations:
  \begin{equation}
c_c(t)=2c_a(t)\qquad\forall t>0\quad ,
  \end{equation}
if initially $c_c(0)=2c_a(0)$. This result has been already derived by
several authors (see for example~\cite{SPOU,PRIV,MALTE}). Moreover, the
relation among the actions ${\cal S}_c$ and ${\cal S}_a$ has been
already used in previous works~\cite{PELITI_UNO,SAS} for Poissonian initial
conditions. However, our results are valid for all the $\alpha$-models and
take care explicitly of arbitrary initial condition. In addition, from
eq.~(\ref{equ:equiv}), we clearly see that quantities involving correlation
functions of the same order are simply related. But more complicated
quantities as for example the interparticle distribution function (which
involves many correlation functions of different order), are no longer
simply related. This explain why the asymptotic interparticle distribution
function between DC and DA model are qualitatively different\cite{BEN}.

\section{The one-dimensional coagulation model revisited}

Let us define ${\cal E}(x,y,t)$, the time dependent probability that the
interval $(y,x)$ is empty at time $t$. Our goal is to rederive in the
framework of the field theory the equation of motion for ${\cal E}(x,y,t)$
obtained by Doering {\it et al}.~\cite{LIN}. This equation reads:
  \begin{equation}
{\partial {\cal E}(x,y,t) \over \partial t} =
D\left( {\partial ^2 \over \partial x^2}+{\partial ^2 \over \partial y^2}
\right) {\cal E}(x,y,t)\quad .
  \label{equ:charlie}\end{equation}
First, we have to identify the Fock space operator corresponding to the
probability ${\cal E}(x,y,t)$. This means that we are looking for a
time-independent operator $E(x,y)$ such that, given a state $\vert \Phi(t)
\rangle$, the probability that the interval between $y$ and $x$
is empty at time $t$ is given by
  \begin{equation}
{\cal E}(x,y,t)=\langle \vert E(x,y)\vert \Phi(t)\rangle \quad .
  \label{equ:defEdue}\end{equation}
This operator is:
  \begin{equation}
E(x,y)=e^{-\int_{y}^{x} dz\psi(z)}\quad .
  \label{equ:defE}\end{equation}
The easiest way to show this consists in using the explicit form of the
projection state
$\langle \vert=\langle 0\vert e^{\int_{-\infty}^{+\infty} dz\psi(z)}$ in
eq. (\ref{equ:defEdue}). Indeed, for a general state $\vert\Phi\rangle$,
we find:
$${\cal E}(x,y)=\langle 0\vert e^{(\int_{x}^{\infty} +\int_{-\infty}^{y})
dz\ \psi(z)}\vert \Phi\rangle\quad .$$
If we consider the state
$\vert \Phi^{\prime}\rangle=\pi(z_1)\dots \pi(z_n)\vert \Phi\rangle$
obtained by adding
particles in  $z_1 ,z_2 ,\dots,z_n$ to the state $\vert\Phi\rangle$
we obtain:
  \begin{equation}
\langle E(x,y)\rangle _{\Phi^{\prime}}=\cases{\langle E(x,y)\rangle_\Phi
&if\ $z_n \not\in (y,x)\ \forall n,$ \cr 0 &otherwise.}
  \label{equ:a}\end{equation}
This defines the probability ${\cal E}(x,y)$ introduced above. Moreover,
we have:
\begin{equation}
\psi(x)=-\left[{\partial \over \partial x}E(x,y)\right]_{x=y}\quad ,
  \label{equ:psi-E}\end{equation}
in agreement with Doering {\it et al}.~\cite{LIN}.

The basic commutation relation is:
  \begin{equation}
[E(x,y),\pi(z)]=-f(z,x,y)E(x,y)\quad ,
  \label{equ:commE}\end{equation}
where the function $f$ is given by:
  \begin{equation}
f(z,x,y)=\cases {1 &if\ $y<z<x,$ \cr 0 &otherwise.}
  \label{equ:f}\end{equation}
This commutation relation can be used to prove eq.~(\ref{equ:a}) without
any reference to the explicit form of the projection state.

We can now derive the equation of motion for ${\cal E}(x,y,t)$.
{}From eqs.~(\ref{equ:H_c},\ref{equ:evoluzioneA}) and the property
(\ref{equ:a}), we have:
  \begin{eqnarray}
{\partial \over \partial t}\langle E(x,y)\rangle&=&
\langle E(x,y)(H_0+H_c)\rangle\label{equ:evol}\\
&=&-D\langle E(x,y)[\psi^{\prime}(x)-\psi^{\prime}(y)]\rangle
+\int_{\Sigma} dzdz^{\prime}V(\vert z-z^{\prime}\vert)
\langle E(x,y)\psi(z)\psi(z^{\prime})\rangle\quad ,
  \nonumber\end{eqnarray}
where the domain of integration is
$$\Sigma=\{ (z,z^{\prime})\ \vert\ z\not\in (y,x),\ z^{\prime}\in (y,x)\}$$
To proceed, we have to specify  the interaction $V(\xi)$.
We choose the following form:
  \begin{equation}
V(\xi)=\cases {v &if $0<\xi<\sigma,$ \cr 0 &if $\xi>\sigma$}
  \label{equ:V}\end{equation}
and  we shall eventually take the $\sigma \rightarrow 0$ limit.
For this purpose let us define:
$$v\sigma =\lambda \quad\hbox{and}\quad v\sigma ^2 =\chi\quad .$$
Taking into account the form of the domain of integration
$\Sigma$, the expansion in $\sigma$ of eq. (\ref{equ:evol}) gives:
  \begin{eqnarray}
{\partial \over \partial t}\langle E(x,y)\rangle&=&
\langle E(x,y)(H_0+H_c)\rangle\label{equ:expansion}\\
&=&-D\langle E(x,y)[\psi^{\prime}(x)-\psi^{\prime}(y)]\rangle
+{1\over 2}\chi \langle E(x)[\psi^2 (x)+\psi^2 (0)]\rangle +O(\sigma ^3 )
  \nonumber\end{eqnarray}
{}From the definition of $E(x,y)$ (eq.~(\ref{equ:defE})), we obtain:
  \begin{equation}
({\partial^2 \over \partial x^2}+{\partial^2 \over \partial y^2})
\langle E(x,y)\rangle
=-\langle E(x,y)[\psi^{\prime}(x)-\psi^{\prime}(y)]\rangle
+\langle E(x,y)[\psi^2 (x)+\psi^2 (y)]\rangle\quad .
  \label{equ:derE}\end{equation}
Thus in the limit
  \begin{equation}
\sigma \rightarrow 0, \qquad v \rightarrow \infty \qquad
\hbox{such that} \qquad v\sigma^2 =\chi \rightarrow 2D\quad ,
  \label{equ:limite}\end{equation}
we recover the desired equation:
\begin{equation}
{\partial \over \partial t}\langle E(x,y)\rangle =
D\left( {\partial ^2 \over \partial x^2}+{\partial ^2 \over \partial y^2}
\right) \langle E(x,y)\rangle +O(\sigma^3)\quad .
\label{equ:charlie2}\end{equation}
The particular choice $\chi\rightarrow 2D$ has its physical motivation in the
relation between a model on a discrete lattice and its continuous
version~\cite{NEXT}. Moreover, our closed equation is valid only in the
limit~(\ref{equ:limite}), which corresponds to an infinite pointlike coupling:
  \begin{equation}
V(\vert z-z^\prime \vert)\rightarrow 2\lambda \delta(z-z^\prime ),
\quad\hbox{with}\quad\lambda=v\sigma \rightarrow \infty\quad .
  \label{equ:lambdainfinito}\end{equation}
This fact is not surprising, considering that such a coupling
corresponds to an instantaneous coagulation on contact: this is exactly the
condition which was used in all the exact
solutions~\cite{RACZ,LUSH,BALD,SPOU,PRIV,GRYN}.

This aspect can also be nicely understood in the renormalization group
framework (see for example~\cite{LEE}). The limit $\lambda\rightarrow\infty$
corresponds to approaching the non-trivial fixed point of the theory and thus
to be in
the asymptotic long time regime. In consequence, we can interpret the
result~(\ref{equ:charlie2}), valid in the limits~(\ref{equ:limite})
and~(\ref{equ:lambdainfinito}), as an equation for the asymptotic long time
regime for a diffusion-coagulation problem, with arbitrary reaction rate.

\section{Generalization  to $\alpha$-reactions}

Following the results obtained in section 2, on the correspondence between
the coagulation model and an $\alpha$-model, we can now find the quantity
corresponding to $E$ for the $\alpha$-model and its equation of motion.

Let us define the operator:
  \begin{equation}
T^{(\alpha)}(x,y)=e^{-\alpha\int_{y}^{x} dz\psi(z)}\quad .
  \label{equ:defR}\end{equation}
Thus, ${\cal T}^{(\alpha)}(x,y)=\langle T^{(\alpha)}(x,y)\rangle$
satisfies the closed equation:
  \begin{equation}
{\partial \over \partial t}{\cal T}^{(\alpha)}(x,y,t)=
D\left( {\partial ^2 \over \partial x^2}+{\partial ^2 \over \partial y^2}
\right) {\cal T}^{(\alpha)}(x,y,t) +O(\sigma^3)\quad .
  \label{equ:charlie3}\end{equation}
To see the physical meaning of this operator, we consider a state
$\vert \Phi\rangle=\pi(z_1)\dots \pi(z_n)\vert 0\rangle$. It is then easy
to show that:
  \begin{equation}
\langle T^{(\alpha)}(x,y)\rangle_{\Phi} =\prod_{i=1}^n a_i\quad\hbox{with}\quad
a_i=\cases {1 &if $z_i\not\in (y,x)$, \cr 1-\alpha &if $z_i\in (y,x)$. \cr}
  \label{equ:interpretazioneT}
\end{equation}
${\cal T}^{(2)}(x,y)$, which is relevant for the annihilation case,
is related to the probability of having an even number of particles
in the interval between $y$ and $x$. For a generic vector of the form
$\vert \Phi\rangle=F[\pi]\vert 0\rangle$, we also have:
  \begin{equation}
{\cal T}^{(\alpha)}(x,y)=F[1-f^{(\alpha)}(z)]\quad ,
  \label{equ:TsuF}\end{equation}
where the function $f^{(\alpha)}$ is the generalization of~(\ref{equ:f}),
{\it i.e.}\/,
$$f^{(\alpha)}(z)=\cases{\alpha &if $y<z<x$, \cr 0& {\rm otherwise.}\cr}$$

Eq.~(\ref{equ:charlie3}) has been solved by Doering {\it et al.}~\cite{BUR},
with appropriate boundary conditions at $x=y$ and $x=\infty$ and the initial
condition ${\cal T}^{(\alpha)}(x,y,0)={\cal T}_0^{(\alpha)}(x,y)$. The
solution of this equation is particularly simple if the system is homogeneous.
{}From translational invariance ${\cal T}^{(\alpha)}(x,y,t)=
{\cal T}^{(\alpha)}(x-y,t)$ and eq.~(\ref{equ:charlie3}) becomes:
  \begin{equation}
{\partial \over \partial t}{\cal T}^{(\alpha)}(x,t) =
2D{\partial ^2 \over \partial x^2}{\cal T}^{(\alpha)}(x,t)\quad ,
  \label{equ:charlieomogenea}\end{equation}
with the conditions:
$${\cal T}^{(\alpha)}(0,t)=1\ , \quad
\vert {\cal T}^{(\alpha)}(\infty,t)\vert \leq 1\quad\hbox{and}\quad
{\cal T}^{(\alpha)}(x,0)={\cal T}^{(\alpha)}_0 (x)\quad .$$
The second condition follows from the fact that $T^{(\alpha)}(x,y)$
is a stochastic variable which takes its value in the
interval $[-1,+1]$.
 To take care of the  first two boundary conditions, it is judicious
to define:
  \begin{equation}
t^{(\alpha)}(x,t) \equiv
{\partial ^2 \over \partial x^2}{\cal T}^{(\alpha)}(x,t)\quad ,
  \label{equ:deft}\end{equation}
whose equation of motion is:
  \begin{equation}
\left\{ \begin{array}{ll}
{\displaystyle \frac{\partial}{\partial t}t^{(\alpha)}(x,t)=
2D{\partial ^2 \over \partial x^2} t^{(\alpha)}(x,t)}\quad ,\\ [.4cm]
t^{(\alpha)}(0,t)=t^{(\alpha)}(\infty,t)=0\quad .
  \end{array}\right.
  \end{equation}
The solution reads:
  \begin{equation}
t^{(\alpha)}(x,t)=\int_0 ^{\infty}dz\ G(x,z,t)t^{(\alpha)}_0 (z)\quad ,
  \label{equ:soluzionet}\end{equation}
where $G(x,z,t)$ is the Green function:
  \begin{equation}
G(x,z,t)=(8\pi Dt)^{-1/2} \left[{\rm exp}\left(-{(x-z)^2 \over 8Dt}\right)
-{\rm exp}\left(-{(x+z)^2 \over 8Dt}\right)\right]\quad .
  \end{equation}

We can now study the behavior of the particle density. From
eq.~(\ref{equ:defR}), it follows that the (homogeneous) density is:
  \begin{equation}
c_{\alpha}(t)=\langle\psi(0)\rangle=-\alpha^{-1}\left[{\partial \over \partial
x}
{\cal T}^{(\alpha)}(x,t) \right]_{x=0}
  \label{equ:densita}\end{equation}
and then from~(\ref{equ:deft}):
  \begin{equation}
c_{\alpha}(t)=\alpha^{-1}\int_0 ^{\infty}dz\ t^{(\alpha)}(z,t)\quad .
  \end{equation}
Using eq.~(\ref{equ:soluzionet})
we obtain:
  \begin{eqnarray}
c_{\alpha}(t)&=&\alpha^{-1}\int_0 ^{\infty}dz\ {\rm erf}
\left({z\over (8Dt)^{1/2}}\right)t_0^{(\alpha)}(z)\\
&=&\alpha^{-1}(2\pi Dt)^{-1/2}\int_0 ^{\infty}d\xi\ e^{-\xi^2}\xi
\left[1-{\cal T}^{(\alpha)}_0(2\xi t^{1/2})\right]\quad ,
  \nonumber\end{eqnarray}
whose asymptotic behavior for large $t$ is:
  \begin{equation}
c_{\alpha}(t)\sim
(2\alpha)^{-1}(2\pi Dt)^{-1/2}\left[1-{\cal T}^{(\alpha)}_0(\infty)\right]\quad
{}.
  \label{equ:densitaasintotica}\end{equation}
This result shows that, besides the trivial dependence on the factor
$(2\alpha)^{-1}$ which follows from the rescaling $\psi\rightarrow\alpha\psi$,
the asymptotic density amplitude depends also on the initial value
${\cal T}^{(\alpha)}_0(\infty)$. The study of this quantity for a general
initial state is a very difficult task. From~(\ref{equ:TsuF})
and~(\ref{equ:state}) we have:
  \begin{equation}
{\cal T}_0^{(\alpha)}(x)={\rm exp}\left[
\sum_{p=1}^{\infty}{(-\alpha)^p \over p!}{\bar g}_{(p)}(x)\right]\quad ,
  \end{equation}
where
$${\bar g}_{(p)}(x)=
\int_{0}^{x}dx_1 \dots \int_{0}^{x}dx_p\
g_{(p)}(x_1 ,\dots,x_p)$$
and it is difficult to predict the value ${\cal T}_0^{(\alpha)}(\infty)$,
without expliciting the $g_{(p)}$'s.

Let us then consider a translational invariant state having the following
properties:
  \begin{equation}
g_{(1)}(x)=c\ ,\quad g_{(2)}(x_1,x_2)=g(\vert x_1-x_2\vert)\quad\hbox{and}
\quad g_{(p)}=0\ ,\ \hbox{for}\ p\geq 3\quad .
  \label{equ:statoparticolare}\end{equation}
Such a state can be easily built, as we shall see later. Thus, we have:
  \begin{equation}
{\cal T}_0^{(\alpha)}(x)={\rm exp}\left\{ -\alpha cx +{\alpha^2 \over 2}
{\bar g}_{(2)}(x)\right\}\quad ,
  \label{equ:tinit}\end{equation}
with
$${\bar g}_{(2)}(x)=2x\int_{0}^{x}dz\ g(z)-2\int_{0}^{x}dz\ z\,g(z)\quad .$$
If we define
$$A=2\int_{0}^{\infty}dz\ g(z)\quad\hbox{and}\quad
B=\int_{0}^{\infty}dz\ z\,g(z)\quad ,$$
which are finite for short range correlations, one sees that
in the limit $x\rightarrow\infty$ three cases have to be distinguished:
  \begin{eqnarray}
1)\ \ \alpha c-{\alpha^2 \over 2}A>0 & : & {\cal T}_0^{(\alpha)}(\infty)=0
    \nonumber \\
2)\ \ \alpha c-{\alpha^2 \over 2}A<0 & : & {\cal T}_0^{(\alpha)}(\infty)=\infty
\label{equ:condit}\\
3)\ \ \alpha c-{\alpha^2 \over 2}A=0 & : & {\cal T}_0^{(\alpha)}(\infty)
    =e^{-\alpha^2B} \nonumber
  \end{eqnarray}
The case $2)$ is unphysical, because the particle density must be positive.
Thus the value of $A$ is such that $\alpha A\leq 2c$ for all $\alpha\in[1,2]$,
which implies $A\leq c$. This constrain is a consequence of the assumption
$g_{(p)}=0$ for $p\geq 3$ and $B<\infty$. This also means that the case $3)$
is only realizable for $\alpha=2$, $A=c$. From~(\ref{equ:densitaasintotica})
it follows that for an initial state of the type~(\ref{equ:statoparticolare}),
we have the same asymptotic density amplitude for all the $\alpha$-models
with $1\leq \alpha <2$ and for the pure-annihilation case with $A<c$. But in
the pure-annihilation case with $A=c$, the amplitude depends on the value of
$B$.

We now show how to built  a state of the form
(\ref{equ:statoparticolare}) and understand its physical content.
Let us take a Poissonian state with density ${\hat c}$:
$$\vert P\rangle =e^{{\hat c}\int dz\ (\pi(z)-1)}\vert 0\rangle$$
and the state $\vert \Phi_{\beta}\rangle$ obtained by the replacement:
  \begin{equation}
\pi(z) \rightarrow \pi(z)\left
[\beta+{1 \over 2{\hat c}}\int dy\ \pi(z+y)g(y)\right]\ \ \ \ \ \
0\leq \beta \leq 1
  \label{equ:b}\end{equation}
The normalization $\langle \vert \Phi_\beta\rangle=1$ implies that:
$$A=2\int_{0}^{\infty}dy\ g(y)=2(1-\beta){\hat c}\quad .$$
The modification (\ref{equ:b}) means that we go from the state
$\vert P\rangle$, where particles
are randomly distributed, to the state $\vert \Phi_{\beta}\rangle$,
where both single particles and pairs are
randomly distributed.
The probability density that a pair have an extension
$\sigma$ is ${1 \over 2{\hat c}}g(\sigma)$.
If $\beta=0$ there are only pairs in the state $\vert \Phi_0\rangle$ .
A straightforward calculation gives:
$$\vert \Phi_{\beta}\rangle={\rm exp}\left\{ (2-\beta){\hat c}\int dz\
(\pi(z)-1)+{1 \over 2}\int dzdy\ g(z-y)(\pi(z)-1)(\pi(y)-1)\right\}\quad .$$
Thus we have a state of the type~(\ref{equ:statoparticolare}) with density
$c=(2-\beta){\hat c}$. For $\beta=0$, we have $A=c$. The mean extension of
a pair is:
$${\bar \sigma}={1\over 2{\hat c}}\int_0^{\infty}d\sigma\ \sigma g(\sigma)=
{(2-\beta)B\over 2c}\quad .$$
Our $\alpha$-models, with initial correlations can now be parametrized in
terms of $(\alpha,\beta,{\bar\sigma})\in [1,2]\times [0,1]\times [0,\infty)$.
The asymptotic density amplitudes are universal for all the values of the
parameters, except on the line $(\alpha=2,\beta=0,{\bar\sigma})$ where the
amplitude is proportional to $(1-e^{-4c{\bar\sigma}})$. In this case, the
asymptotic density amplitude tends to zero if ${\bar \sigma}\rightarrow 0$.
The physical reason is that when ${\bar \sigma}=0$, the two particles of a pair
are at the same place and immediately annihilate. This situation cannot occur
if
coagulation is present, because single isolated particles will remain.

It would be interesting to see how a more general initial state would affect
the amplitudes. Note finally, that the relations~(\ref{equ:condit}),
characterizing the initial state, are preserved by the dynamical evolution.
Indeed, every evolved state can be considered as a new initial state evolving
towards the same asymptotic solution.

\vskip 1truecm
\section{Conclusions}

We have studied a class of $d$-dimensional diffusion-reaction models
interpolating continuously between the diffusion-coagulation and the
diffusion-annihilation models. The field theoretical approach used leads
to exact relations between the observables of these different models. In one
dimension, it was shown how correlations in the initial state can lead to a
violation of the universality of the amplitude for the DC models.

Several extension of the present work concerning one-dimensional models with
reversible diffusion-reaction systems, or the presence of fronts in
inhomogeneous systems are under investigation. Moreover, one may expect
to construct ``good observables'' in $d$-dimensional systems for which a
closed equation of motion can be derived.

\section*{Acknowledgements}

We thanks Z. R\'acz for many useful discussions.

\end{document}